\documentclass[letterpaper,english,reprint, aps]{revtex4-1}
\usepackage[T1]{fontenc}
\usepackage[latin9]{inputenc}
\usepackage{color}
\usepackage{amsmath}
\usepackage{amssymb}
\usepackage{graphicx}
\usepackage{setspace}

\makeatletter

\newcommand*\LyXThinSpace{\,\hspace{0pt}}
\pdfpageheight\paperheight
\pdfpagewidth\paperwidth

\makeatother

\usepackage{babel}
\begin{document}
\title{Coherence induced work in quantum heat engines with Larmor precession}
\author{Shanhe Su}
\address{Department of Physics, Xiamen University, Xiamen, 361005, China}
\author{Wei Shen}
\address{Department of Physics, Xiamen University, Xiamen, 361005, China}
\author{Ran Liu$^{1\text{,2,3}}$}
\address{$^{1}$Hefei National Laboratory for Physical Sciences at the Microscale
and Department of Modern Physics, University of Science and Technology
of China, Hefei, 230026, China~\\
$^{2}$CAS Key Laboratory of Microscale Magnetic Resonance, University
of Science and Technology of China, Hefei, 230026, China~\\
$^{3}$Synergetic Innovation Center of Quantum Information and Quantum
Physics, University of Science and Technology of China, Hefei, 230026,
China}
\author{Min Jiang$^{1\text{,2,3}}$}
\address{$^{1}$Hefei National Laboratory for Physical Sciences at the Microscale
and Department of Modern Physics, University of Science and Technology
of China, Hefei, 230026, China~\\
$^{2}$CAS Key Laboratory of Microscale Magnetic Resonance, University
of Science and Technology of China, Hefei, 230026, China~\\
$^{3}$Synergetic Innovation Center of Quantum Information and Quantum
Physics, University of Science and Technology of China, Hefei, 230026,
China}
\author{Ji Bian$^{1\text{,2,3}}$}
\address{$^{1}$Hefei National Laboratory for Physical Sciences at the Microscale
and Department of Modern Physics, University of Science and Technology
of China, Hefei, 230026, China~\\
$^{2}$CAS Key Laboratory of Microscale Magnetic Resonance, University
of Science and Technology of China, Hefei, 230026, China~\\
$^{3}$Synergetic Innovation Center of Quantum Information and Quantum
Physics, University of Science and Technology of China, Hefei, 230026,
China}
\author{Xinhua Peng$^{1\text{,2,3}}$}
\email{xhpeng@ustc.edu.cn}

\address{$^{1}$Hefei National Laboratory for Physical Sciences at the Microscale
and Department of Modern Physics, University of Science and Technology
of China, Hefei, 230026, China~\\
$^{2}$CAS Key Laboratory of Microscale Magnetic Resonance, University
of Science and Technology of China, Hefei, 230026, China~\\
$^{3}$Synergetic Innovation Center of Quantum Information and Quantum
Physics, University of Science and Technology of China, Hefei, 230026,
China}
\author{Jincan Chen}
\email{jcchen@xmu.edu.cn}

\address{Department of Physics, Xiamen University, Xiamen, 361005, China}
\begin{abstract}
\singlespacing{}The impacts of quantum coherence on nonequilibrium thermodynamics
become observable by dividing the heat and work into the conventional
diagonal part and the other part relaying on the superpositions and
the time derivative of Hamiltonian. Specializing to exactly-solvable
dynamics of Larmor precession, we build a quantum Otto heat engine
employing magnetic-driven atomic rotations. \textcolor{black}{The
coherence induced by the population transition guarantees the positive
work output when the control protocol is time dependent. }The time-dependent
control of a quantum heat engine implements the correspondence between
the classical and quantum adiabatic theorems for microscopic heat
machines.
\end{abstract}
\maketitle

\section{INTRODUCTION}

The first law of thermodynamics in any infinitesimal process can be
expressed by taking the differential of the internal energy \citep{key-1,key-2}.
For an open quantum system, the heat was defined originally by Alicki
\citep{key-3,key-4} as the non-unitary dissipative energy exchange
due to the interaction between the system and the bath, while the
work was described by the time-varying of Hamiltonian. Based on the
first law of thermodynamics in the quantum domain, Kosloff et al.
first did a systematic study of quantum heat engine (QHE) cycles working
with harmonic oscillators and spins \citep{key-5,key-6,key-7,key-8}.
Boukobza et al. extended Alicki\textquoteright s formulas into the
Heisenberg and interaction pictures and illustrated thermodynamics
of bipartite systems \citep{key-9,key-10}. Quantum coherence may
stimulate additional energy changes in thermodynamic processes, which
has been observed by expressing the heat and work in term of the instantaneous
orthonormal basis \citep{key-11,key-12}. 

Nowadays, quantum thermodynamics has aroused general interest in both
research and practice. Numerous unique properties arise due to quantum
effects in the operation of microscopic heat engines. Klatzow et al.
used nitrogen vacancy centers in the diamond to implement a three-level
engine with long-lived coherence at the room temperature \citep{key-13}.
Pekola et al. realised the miniature Otto cycle by exploiting the
time-domain dynamics and coherence of driven superconducting qubits
\citep{key-14,key-15,key-16}. A quantum Otto heat engine (QOHE) operating
under the reservoir at effective negative temperature was experimentally
performed by employing Carbon-13 NMR spectroscopy \citep{key-17,key-18}.
Quantum machines powered by nonthermal energy sources such as externally
injected coherent atoms \citep{key-19,key-20,key-21} or squeezed
baths \citep{key-22,key-23,key-24} have been shown to exhibit unconventional
performances. Quantum optomechanical realization of a heat engine
generates alternative strategies to extract work from the thermal
energy of a mechanical resonator \citep{key-25,key-26,key-27}. Other
proposals may focus on QHEs when the fluctuation relation enters the
conventional trade-off between the power and the efficiency \citep{key-28,key-29,key-30,key-31}. 

However, a QOHE undergoes a four-step cycle where the two adiabatic
branches involve the quantum adiabatic approximation. It means that
a physical system should carry out a slow down evolution with a time-independent
Hamiltonian ($\frac{\partial H}{\partial t}\rightarrow0$) and remain
in the instantaneous eigenstate corresponding to the initial Hamiltonian
\citep{key-32,key-33,key-34}. QOHEs based on quantum adiabatic processes
are insufficient to incorporate quantum effects into the performance
evaluation of a heat engine. Given this context, we are naturally
led to consider the thermodynamic cycle applying time-varying Hamiltonians
in the adiabatic processes. The thermodynamic quantities expressed
in terms of the instantaneous eigenvectors of Hamiltonian allow estimating
the capabilities of engines energised by quantum coherence.

In this work, based on the definitions of heat and work with respect
to generic time-dependent open systems, a four-stroke power cycle,
followed by thermodynamic adiabatic compression and expansion, and
isochoric heat input and output, is built. We consider an atom driven
by a rotating magnetic field as the working substance and give the
formula of the work done when the atom experiences adiabatic evolution
from an equilibrium state. How the quantum coherence induced by the
transitions between different eigenstates influences the performance
of a heat engine is waiting to be discovered. Importantly, we are
interested in the correspondence between the classical and quantum
adiabatic regimes for QHEs.

\section{HEAT AND WORK IN QUANTUM THERMODYNAMIC PROCESSES }

To build a quantum heat engine, a preliminary and necessary step is
to identify the heat and work in quantum regimes. In our previous
study \citep{key-11}, heat and work are classified on the basis of
the time-dependent processes. According to the microscopic description
of the first law of thermodynamics, the rates of the heat $\dot{Q}$
absorbed from the surroundings and the work $\dot{W}$ performed by
the external field have the following forms

\begin{equation}
\dot{Q}=\sum_{n}\dot{\rho}_{nn}E_{n}-\sum_{n\neq m}\rho_{nm}\left\langle m\right|\frac{\partial\hat{H}}{\partial t}\left|n\right\rangle 
\end{equation}
and

\begin{equation}
\dot{W}=\sum_{n}\rho_{nn}\dot{E}_{n}+\sum_{n\neq m}\rho_{nm}\left\langle m\right|\frac{\partial\hat{H}}{\partial t}\left|n\right\rangle ,
\end{equation}
where $\left|m\left(t\right)\right\rangle $ denotes the instantaneous
eigenstate of the Hamiltonian $\hat{H}(t)$ with the energy level
$E_{m}\left(t\right)$, $\rho_{nm}\left(t\right)=\left\langle n\left(t\right)\right|\hat{\rho}\left|m\left(t\right)\right\rangle $
represents the element of the density matrix, and the dot indicates
the time derivative. For simplicity purposes, the letter ``(t)''
is omitted in Eqs. (1) and (2). It is worthy of note that Eq. (2)
was also obtained in Ref. \citep{key-12}.

When the quantum system is coupled to a heat bath and the Hamiltonian
is time independent, i.e., $\frac{\partial\hat{H}}{\partial t}=0$
and $\dot{E}_{n}=0$, there is no work done by the external force
$\left(\dot{W}=0\right)$. This process is considered to be isochoric
exemplified by the heating or cooling of the working substance with
$\dot{Q}=\sum_{n}\dot{\rho}_{nn}E_{n}$.

An adiabatic process occurs without the transfer of heat or mass between
the system and the environment. Therefore, the alteration of the internal
energy exists in the form of work only. The Liouville-von Neumann
equation describes how the density operator evolves in time \citep{key-35},
i.e.,
\begin{equation}
\dot{\hat{\rho}}=-\frac{i}{\hbar}\left[\hat{H},\hat{\rho}\right].
\end{equation}
Replacing the operator in a matrix form, we are capable of reaching
the equality
\begin{equation}
\sum_{n}\dot{\rho}_{nn}E_{n}=\sum_{n\neq m}\rho_{nm}\left\langle m\right|\frac{\partial\hat{H}}{\partial t}\left|n\right\rangle ,
\end{equation}
which indicates that $\dot{Q}=0$. The quantum coherence, represented
by $\sum_{n\neq m}\rho_{nm}\left\langle m\right|\frac{\partial\hat{H}}{\partial t}\left|n\right\rangle $,
eliminates the heat loss to the surroundings in the thermodynamic
adiabatic process. 

Equations (1) and (2) give the general formulas of the heat and work
in quantum thermodynamic processes, and are compatible with Alicki
and Kieu's definitions \citep{key-3,key-36}. The first term in Eq.
(1) states that the change of the probabilities $\rho_{nn}$ generates
the heat transfer. Work can be done by a system during a process that
alters the energy levels $E_{n}$, as indicated by the first quantity
of Eq. (2). The second parts in Eqs. (1) and (2) imply that $\dot{Q}$
and $\dot{W}$ are closely related to the quantum coherence if the
Hamiltonian depends sensitively on time. In view of the expressions
Eqs. (1) and (2), we are going to build a Otto quantum heat engine
relying on a time-dependent adiabatic process {[}Fig. 1(a){]}.

\begin{figure}
\includegraphics[scale=0.35]{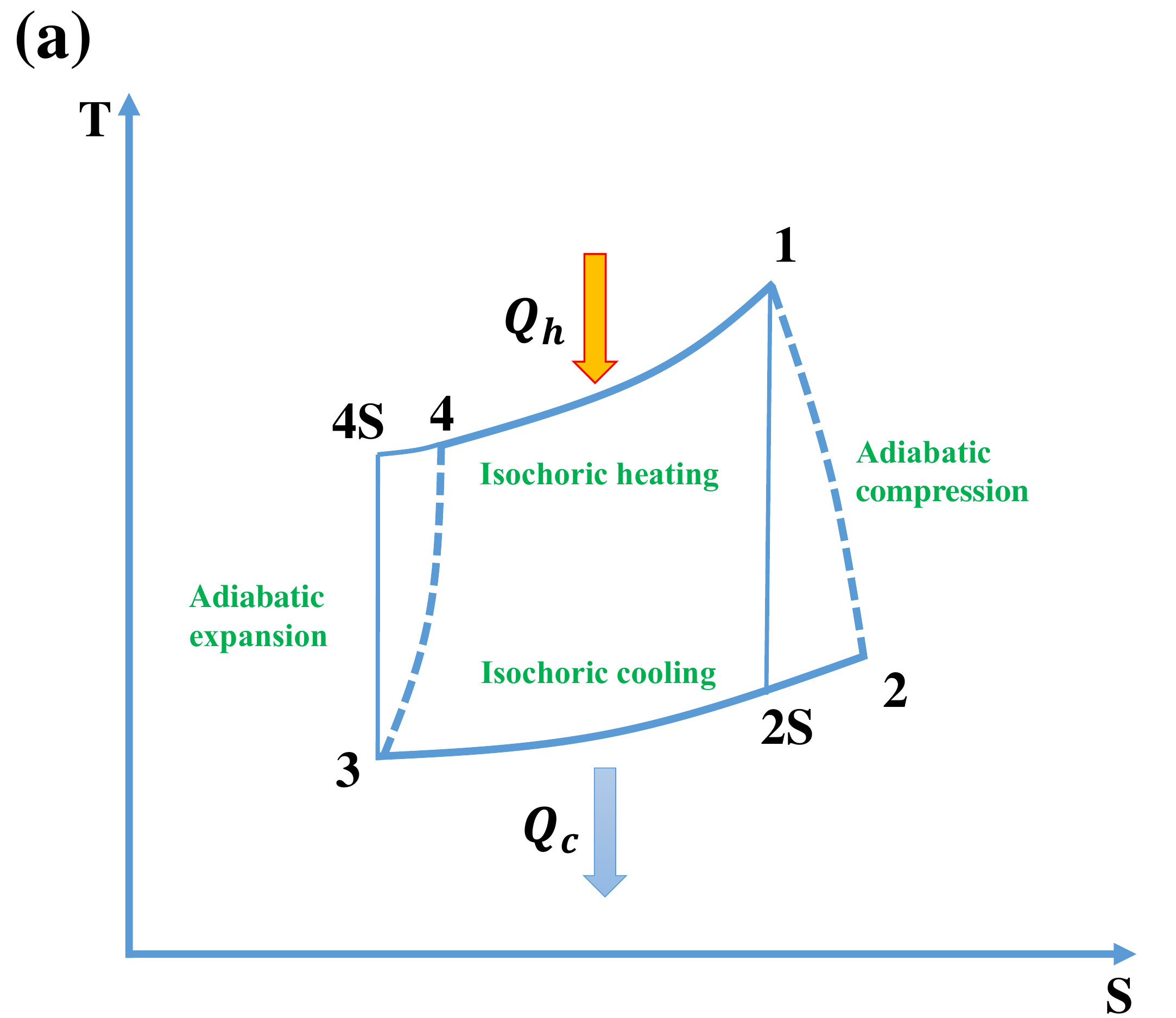}

\includegraphics[scale=0.23]{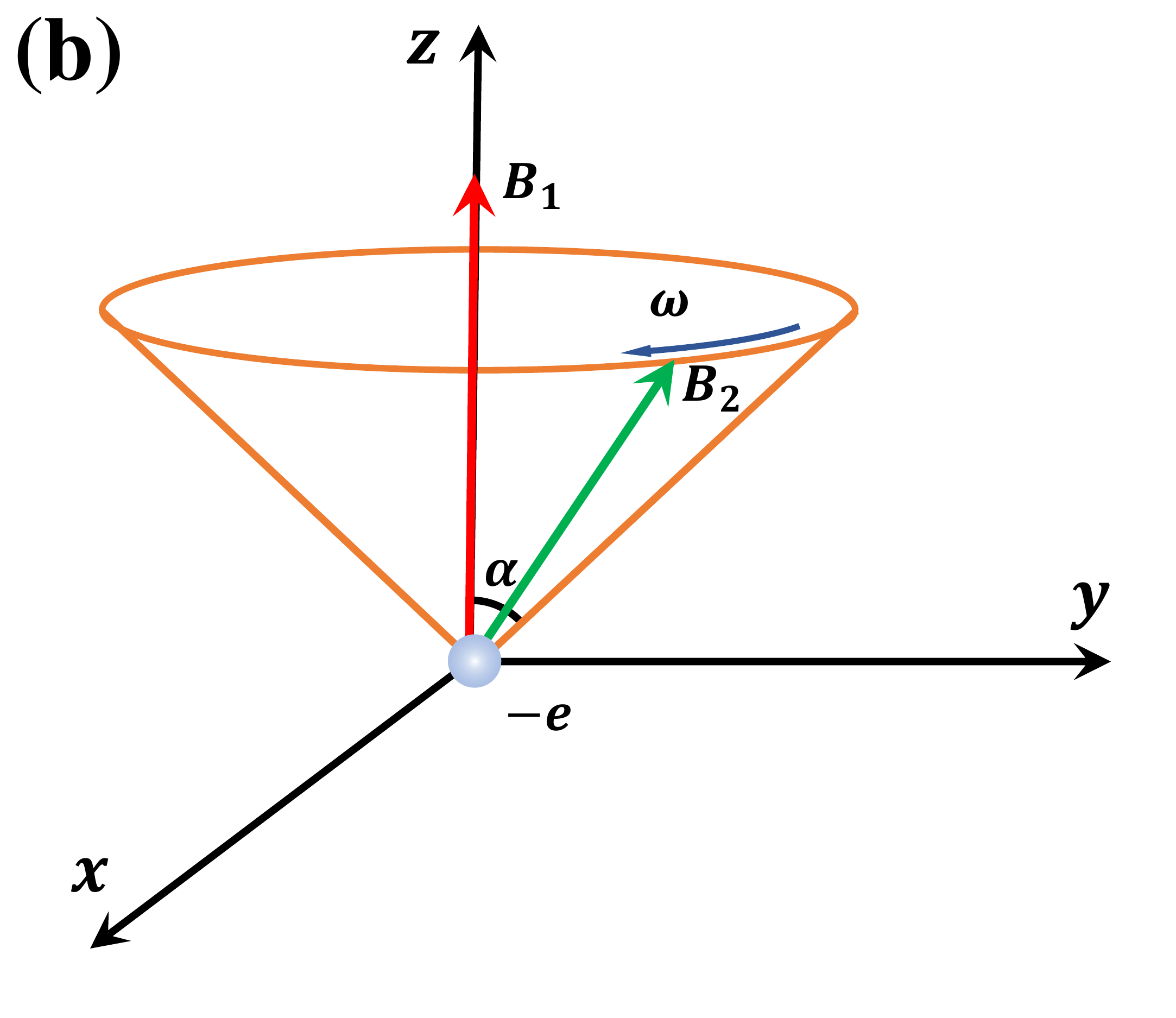}\includegraphics[scale=0.23]{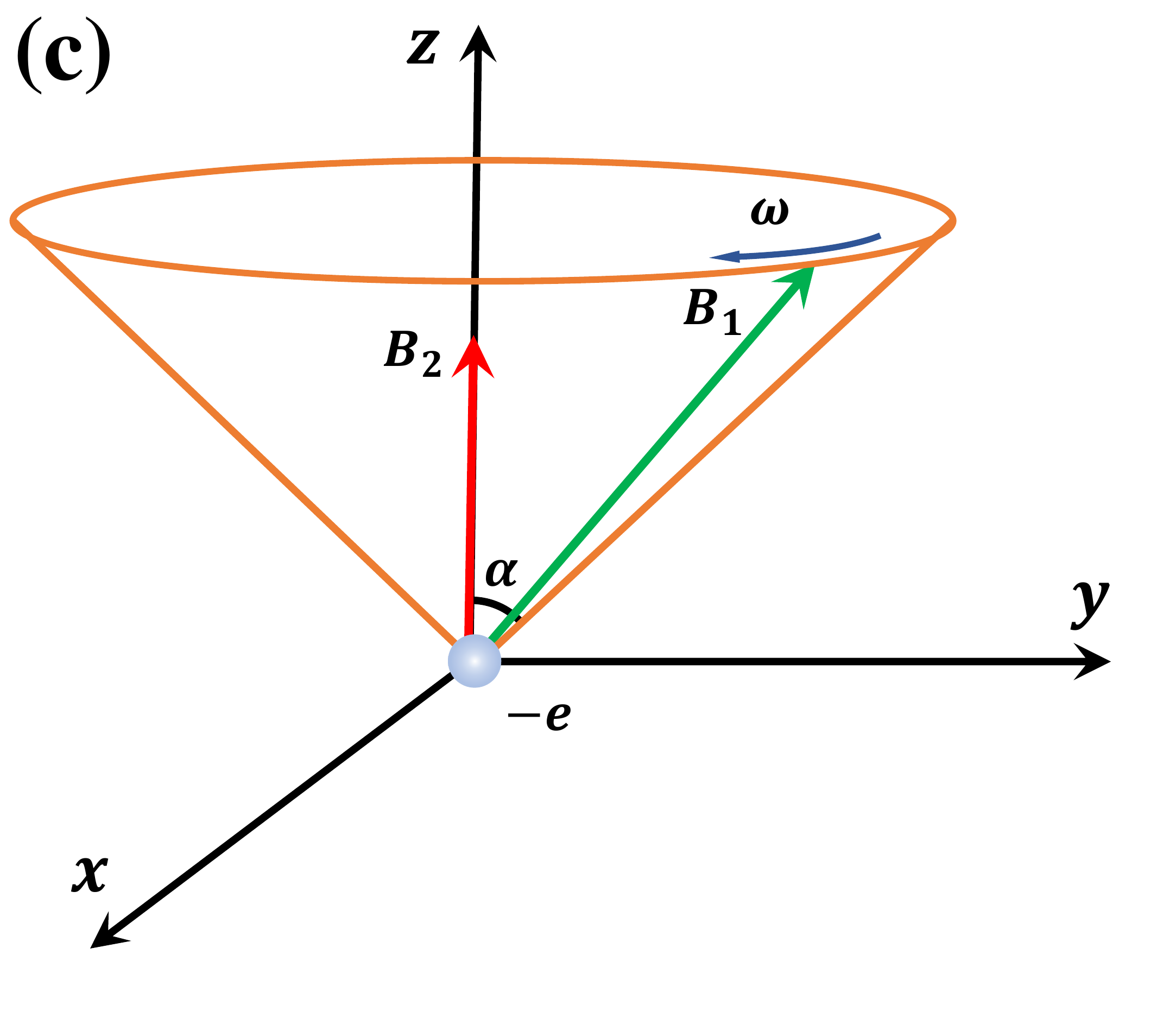}

\caption{(a) The temperature\textendash entropy (T\textendash S) diagram of
the quantum Otto heat engine based on the atomic Larmor precession.
The process from 1 to 2 (3 to 4) represents the irreversible adiabatic
compression (expansion) process, where the magnetic field {[}green
arrows in (b) and (c){]} starts to ride around at the intensity $B_{2}\left(B_{1}\right)$,
angular velocity $\omega$, and incline angle $\alpha$. The process
from 4 to 1 (2 to 3) denotes the isochoric heating (cooling) process,
in which the system interacting with a time-independent magnetic field
$\boldsymbol{B}=B_{1}\boldsymbol{e}_{z}\left(B_{2}\boldsymbol{e}_{z}\right)$
{[}red arrows in (b) and (c){]} is put in thermal contact with a hot
(cold) bath at the inverse temperature $\beta_{h}$ $\left(\beta_{c}\right)$.
The evolutions from 1 to 2S and from 3 to 4S represent the quantum
adiabatic processes with $\alpha=0$.}
\end{figure}

\section{THE ADIABATIC EVOLUTION OF THE ATOMIC SYSTEM FROM A STATE OF THERMAL
EQUILIBRIUM}

Before building a four-stroke cycle, one needs to understand the adiabatic
dynamics of the working substance starting from the equilibrium state.
The atomic Larmor precession allows us to extract work by considering
an atom located at the origin of three-dimensional space and driven
by a rotating magnetic field $\boldsymbol{B}\left(j,\alpha,t\right)=B_{j}[\sin\alpha\cos\left(\omega t\right)\boldsymbol{e}_{x}+\sin\alpha\sin\left(\omega t\right)\boldsymbol{e}_{y}+\cos\alpha\boldsymbol{e}_{z}]$
{[}$j=1$ and $2$, as shown in Figs. 1(b) and 1(c){]}. Its Hamiltonian
takes the form \citep{key-37}

\begin{align}
\hat{H}\left(\omega_{j},\alpha,t\right) & =-\hat{\mathbf{\mathbf{\boldsymbol{\mu}}}}\cdot\boldsymbol{B}\left(j,\alpha,t\right)=\frac{\hbar\omega_{j}}{2}\nonumber \\
 & \times[\sin\alpha\cos\left(\omega t\right)\hat{\sigma}_{x}+\sin\alpha\sin\left(\omega t\right)\hat{\sigma}_{y}+\cos\alpha\hat{\sigma}_{z}],
\end{align}
where the atom has the mass $m$ and charge $-e$; $\mathbf{\hat{\mathbf{\mathbf{\boldsymbol{\mu}}}}}=\gamma_{e}\hat{\boldsymbol{S}}$
is the dipole moment determined by the gyromagnetic ratio $\gamma_{e}=-e/m$
and spin angular momentum $\hat{\boldsymbol{S}}$; $\hat{\sigma}_{i}$
($i=x$, $y$, and $z$) define the Pauli spin matrices; $\omega_{j}=eB_{j}/m$;
and $\hbar$ equals the Planck constant divided by $2\pi$.

From the orthonormal basis $\left|\uparrow\right\rangle =\left(\begin{array}{c}
1\\
0
\end{array}\right)$ and $\left|\downarrow\right\rangle =\left(\begin{array}{c}
0\\
1
\end{array}\right)$ , we can write out the normalized eigenstates of $\hat{H}\left(\omega_{j},\alpha,t\right)$
as
\begin{equation}
\left|\chi_{+}\left(\alpha,t\right)\right\rangle =\cos\frac{\alpha}{2}\left|\uparrow\right\rangle +e^{i\omega t}\sin\frac{\alpha}{2}\left|\downarrow\right\rangle 
\end{equation}
and

\begin{equation}
\left|\chi_{-}\left(\alpha,t\right)\right\rangle =e^{-i\omega t}\sin\frac{\alpha}{2}\left|\uparrow\right\rangle -\cos\frac{\alpha}{2}\left|\downarrow\right\rangle ,
\end{equation}
which represent the spin up and down, respectively, along the instantaneous
direction of $\boldsymbol{B}\left(t\right)$. The corresponding eigenvalues
are 
\begin{equation}
E_{\pm}\left(\omega_{j}\right)=\pm\frac{\hbar\omega_{j}}{2},
\end{equation}
remaining unvarying over time.

By invoking the transformation $\hat{R}_{z}\left(t\right)=e^{i\omega t\hat{\sigma}_{z}/2}$,
Eq. (8) reduces to 
\begin{align}
\hat{H}_{R} & \left(\omega_{j},\alpha\right)=\hat{R}_{z}\left(t\right)\hat{H}\left(\omega_{j},\alpha,t\right)\hat{R^{\dagger}}_{z}\left(t\right)-i\hbar\hat{R}_{z}\left(t\right)\frac{\partial}{\partial t}\hat{R^{\dagger}}_{z}\left(t\right)\nonumber \\
 & =\frac{\hbar}{2}\left[\omega_{j}\sin\alpha\hat{\sigma}_{x}+\left(\omega_{j}\cos\alpha-\omega\right)\hat{\sigma}_{z}\right].
\end{align}
In the rotating frame, the effective Hamiltonian $\hat{H}_{R}\left(\omega_{j},\alpha\right)$
contains no explicit time dependence, yielding the evolution operator

\begin{align}
\hat{U_{R}}\left(\omega_{j},\alpha,t\right) & =e^{-i\hat{H}_{R}\left(\omega_{j},\alpha\right)t/\hbar}\nonumber \\
 & =\cos\varOmega_{j}t/2-i\frac{\omega_{j}\sin\alpha\sin\varOmega_{j}t/2}{\varOmega_{j}}\hat{\sigma}_{x}\nonumber \\
 & -i\frac{\left(\omega_{j}\cos\alpha-\omega\right)\sin\varOmega_{j}t/2}{\varOmega_{j}}\hat{\sigma}_{z}.
\end{align}
In Eq. (10), $\varOmega_{j}=\sqrt{\left(\omega_{j}\cos\alpha-\omega\right)^{2}+\omega_{j}^{2}\sin^{2}\alpha}$,
and the relation $e^{i\mathbf{\hat{\sigma}}\cdot\mathbf{A}}=\cos A+i\frac{\mathbf{\hat{\sigma}}\cdot\mathbf{A}}{A}\sin A$
{[}$\mathbf{\hat{\sigma}}=\hat{\sigma}_{x}+\hat{\sigma}_{y}+\hat{\sigma}_{z}$,
and $A=\left|\mathbf{A}\right|$ {]} has been applied. The evolution
operator in the original frame reads

\begin{align}
\hat{U}\left(\omega_{j},\alpha,t\right) & =\hat{R^{\dagger}}_{z}\left(t\right)\hat{U_{R}}\left(\omega_{j},\alpha,t\right)=\left(\begin{array}{cc}
e^{-i\omega t/2} & 0\\
0 & e^{i\omega t/2}
\end{array}\right)\nonumber \\
 & \times\left(\begin{array}{c}
\cos\varOmega_{j}t/2-i\frac{\omega_{j}\cos\alpha-\omega}{\varOmega_{j}}\sin\varOmega_{j}t/2\\
-i\frac{\omega_{j}\sin\alpha}{\varOmega_{j}}\sin\varOmega_{j}t/2
\end{array}\right.\nonumber \\
 & \left.\begin{array}{c}
-i\frac{\omega_{j}\sin\alpha}{\varOmega_{j}}\sin\varOmega_{j}t/2\\
\cos\varOmega_{j}t/2+i\frac{\omega_{j}\cos\alpha-\omega}{\varOmega_{j}}\sin\varOmega_{j}t/2
\end{array}\right).
\end{align}
The unitary operator immediately allows us to understand the dynamic
behaviors of the atom. 

As an illustrative example, we consider the system initially in a
state of thermal equilibrium (characterized by the ambient temperature
$T$) subjected to a uniform magnetic field in the z-direction $\boldsymbol{B}\left(1,0,0\right)=B_{1}\boldsymbol{e}_{z}$
{[}red arrow in Fig. 1(b){]}. Because $\hat{H}\left(\omega_{1},0,0\right)$
and $\hat{\sigma}_{z}$ commute, this canonical ensemble creates a
diagonal density matrix in the $\hat{\sigma}_{z}$ basis, i.e., 
\begin{equation}
\hat{\rho}_{th}\left(\omega_{1},\beta\right)=\frac{\left(\begin{array}{cc}
e^{-\beta\hbar\omega_{1}/2} & 0\\
0 & e^{\beta\hbar\omega_{1}/2}
\end{array}\right)}{Z\left(\omega_{1},\beta\right)}.
\end{equation}
The partition function $Z\left(\omega_{1},\beta\right)=e^{-\beta\hbar\omega_{1}/2}+e^{\beta\hbar\omega_{1}/2}$.
The inverse temperature $\beta=1/\left(k_{B}T\right)$, where $T$
is the effective temperature and $k_{B}$ is the Boltzmann constant. 

For $t>0$, the magnetic field {[}green arrow in Fig. 1 (b){]} makes
a rotation around the z-axis, given by the vector $\boldsymbol{B}\left(2,\alpha,t\right)$.
The time evolution of the density operator is obtained by the unitary
transformation\textcolor{black}{
\begin{equation}
\hat{\rho}\left(\omega_{2},\omega_{1},\alpha,\beta,t\right)=\hat{S}\left(\alpha,t\right)\hat{\rho}^{\ast}\left(\omega_{2},\omega_{1},\alpha,\beta,t\right)\hat{S}^{\dagger}\left(\alpha,t\right)
\end{equation}
}By defining the transformation matrix $\hat{S}\left(\alpha,t\right)=\left(\begin{array}{cc}
\cos\frac{\alpha}{2} & e^{-i\omega t}\sin\frac{\alpha}{2}\\
e^{i\omega t}\sin\frac{\alpha}{2} & -\cos\frac{\alpha}{2}
\end{array}\right)$, the matrix elements of\textcolor{black}{{} $\hat{\rho}^{\ast}\left(\omega_{2},\omega_{1},\alpha,\beta,t\right)=\hat{U}\left(\omega_{2},\alpha,t\right)\rho_{th}\left(\omega_{1},\beta\right)\hat{U}^{\dagger}\left(\omega_{2},\alpha,t\right)$}
have been written in terms of the eigenstates of the instantaneous
Hamiltonian, which are, respectively, given by $\rho_{nm}\left(\omega_{2},\omega_{1},\alpha,\beta,t\right)=\left\langle \chi_{n}\left(\alpha,t\right)\right|\hat{\rho}\left(\omega_{2},\omega_{1},\alpha,\beta,t\right)\left|\chi_{m}\left(\alpha,t\right)\right\rangle $
($\left\{ n,m\right\} =\left\{ +,+\right\} $, $\left\{ -,-\right\} $,
$\left\{ +,-\right\} $, or $\left\{ -,+\right\} $). 

Most existing literatures studied the quantum thermodynamics based
on the time-independent Hamiltonian or the quantum adiabatic theorem.
Thus, Eqs. (1) and (2) reduce to $\dot{Q}=\sum_{n}\dot{\rho}_{nn}E_{n}$
and $\dot{W}=\sum_{n}\rho_{nn}\dot{E}_{n}$, meaning that the heat
exchange merely depends on the population alteration and the work
corresponds to the change of the eigenenergy spectrum \citep{key-38,key-39,key-40}.
However, for the two-level system driven by a rotating magnetic field,
the eigenvalues relating to the instantaneous eigenstates $\left|\chi_{+}\left(\alpha,t\right)\right\rangle $
and $\left|\chi_{-}\left(\alpha,t\right)\right\rangle $ are fixed
values {[}Eq. (8){]}. If the work flux in the thermodynamic adiabatic
process remains being computed by $\dot{W}=\sum_{n}\rho_{nn}\left(\omega_{2},\omega_{1},\alpha,\beta,t\right)\dot{E}_{n}\left(\omega_{2}\right)$,
one has $\dot{W}=0$. It is anormal that no work can be done by the
rotating magnetic field $\boldsymbol{B}\left(2,\alpha,t\right)$.
In addition, the time derivative of $\rho_{++}\left(\omega_{2},\omega_{1},\alpha,\beta,t\right)$
and $\rho_{--}\left(\omega_{2},\omega_{1},\alpha,\beta,t\right)$
arrives at $\dot{Q}=\dot{\rho}_{++}\left(\omega_{2},\omega_{1},\alpha,\beta,t\right)E_{+}\left(\omega_{2}\right)+\dot{\rho}_{--}\left(\omega_{2},\omega_{1},\alpha,\beta,t\right)E_{-}\left(\omega_{2}\right)=\frac{\hbar\omega\omega_{2}^{2}\sin\varOmega_{2}t\sin^{2}\alpha\tanh\left(\frac{\beta\hbar\omega_{1}}{2}\right)}{2\varOmega_{2}}$.
As $\dot{Q}$ is a non-zero value, it appears unconvincing that the
heat transfer between the system and the environment exists in the
thermodynamic adiabatic process. 

For these reasons, the second terms on the right of Eqs. (1) and (2)
quantifying quantum coherence play an indispensable role in the formulation
of the first law of thermodynamics. Making use of Eq. (5) and taking
the off-diagonal elements of the density operator in Eq. (13), one
readily gets 

\begin{align}
\sum_{n\neq m}\rho_{nm}\left(\omega_{2},\omega_{1},\alpha,\beta,t\right)\left\langle m\right|\frac{\partial\hat{H}\left(\omega_{2},\alpha,t\right)}{\partial t}\left|n\right\rangle \nonumber \\
=\frac{\hbar\omega\omega_{2}^{2}\sin\varOmega_{2}t\sin^{2}\alpha\tanh\left(\frac{\beta\hbar\omega_{1}}{2}\right)}{2\varOmega_{2}},
\end{align}
which satisfies $\sum_{n}\dot{\rho}_{nn}\left(\omega_{2},\omega_{1},\alpha,\beta,t\right)E_{n}\left(\omega_{2}\right)=\sum_{n\neq m}\rho_{nm}\left(\omega_{2},\omega_{1},\alpha,\beta,t\right)\left\langle m\right|\frac{\partial\hat{H}\left(\omega_{2},\alpha,t\right)}{\partial t}\left|n\right\rangle $
and the nonexistence of the heat transfer, i.e., $\dot{Q}=0$. As
a result, the atomic Larmor precession could be regarded as a thermodynamic
adiabatic process. \textcolor{black}{We use 
\begin{align}
W_{L}\left(\omega_{2},\omega_{1},\alpha,\beta,t\right) & =\int_{0}^{t}\sum_{n\neq m}\rho_{nm}\left(\omega_{2},\omega_{1},\alpha,\beta,t^{\prime}\right)\nonumber \\
 & \times\left\langle m\right|\frac{\partial\hat{H}\left(\omega_{2},\alpha,t^{\prime}\right)}{\partial t^{\prime}}\left|n\right\rangle dt^{\prime}\nonumber \\
 & =\frac{\hbar\omega\omega_{2}^{2}\sin^{2}\varOmega_{2}t/2\sin^{2}\alpha\tanh\left(\frac{\beta\hbar\omega_{1}}{2}\right)}{\varOmega_{2}^{2}}
\end{align}
to describe the work performed on the system due to the Larmor precession.
It can be regarded as a coherence work }induced by the transition
between the instantaneous eigenstates $\left|\chi_{\pm}\left(\alpha,t\right)\right\rangle $
of the Hamiltonian $\hat{H}\left(\omega_{2},\alpha,t\right)$. The
work performed by the external field beginning with the initial equilibrium
state is calculated as

\begin{equation}
W\left(\omega_{2},\omega_{1},\alpha,\beta,t\right)=W_{L}\left(\omega_{2},\omega_{1},\alpha,\beta,t\right)+W_{S}\left(\omega_{2},\omega_{1},\alpha,\beta,t\right),
\end{equation}
\textcolor{red}{}where\textcolor{red}{{} }\textcolor{black}{$W_{S}\left(\omega_{2},\omega_{1},\alpha,\beta,t\right)=Tr\left(\left[\hat{H}\left(\omega_{2},\alpha,0\right)-\hat{H}\left(\omega_{1},0,0\right)\right]\hat{\rho}_{th}\left(\omega_{1},\beta\right)\right)+Tr\left(\left[\hat{H}\left(\omega_{2},0,t\right)-\hat{H}\left(\omega_{2},\alpha,t\right)\right]\hat{\rho}^{\ast}\left(\omega_{2},\omega_{1},\alpha,\beta,t\right)\right)$
includes the work contributed by Hamiltonian's sudden shifts from
$\hat{H}\left(\omega_{1},0,0\right)$ to $\hat{H}\left(\omega_{2},\alpha,0\right)$
and from $\hat{H}\left(\omega_{2},\alpha,t\right)$ to $\hat{H}\left(\omega_{2},0,t\right)$. }

\section{QUANTUM OTTO CYCLE}

The quantum Otto cycle is composed of the thermodynamic adiabatic
compression, rejection of heat at a constant external field, thermodynamic
adiabatic expansion, and heat addition at another constant external
field, as illustrated in Fig. 1 (a). The respective scheme of the
four distinct strokes is described below. 

At stage I (1-2), the atom is initially in thermal equilibrium state
$\hat{\rho}_{1}=\hat{\rho}_{th}\left(\omega_{1},\beta_{h}\right)$
characterized by temperature $T_{h}=1K$. From time $t=0$ to $t=\tau_{1}$,
the atom becomes isolated from the hot bath and experiences a thermodynamic
adiabatic compression. The magnetic field is switched from $\boldsymbol{B}\left(1,0,0\right)$
{[}red arrow in Fig. 1(b){]} to $\boldsymbol{B}\left(2,\alpha,0\right)$
{[}green arrow in Fig. 1(b){]} and free to rotate around the z axis.
The working medium unitarily evolves to the mixed state given by $\hat{\rho}_{2}=\hat{\rho}\left(\omega_{2},\omega_{1},\alpha,\beta_{h},\tau_{1}\right)$
{[}see Eq. (13){]}.\textcolor{black}{{} After the rotation, the magnetic
field flips back to the direction of the z axis and remains unvarying
over time, denoted by $\boldsymbol{B}\left(2,0,\tau_{1}\right)$.
}The work performed by the magnetic field increases the internal energy
of the atom, that is,\textcolor{black}{
\begin{equation}
W_{1}=W\left(\omega_{2},\omega_{1},\alpha,\beta_{h},\tau_{1}\right).
\end{equation}
}

At stage II (2-3), the atom having probability of each eigenstate
uniquely determined by $\hat{\rho}_{2}$ comes into contact with the
cold bath at temperature $T_{c}=0.1K$. During the process of reaching
thermal equilibrium, the removal of heat allows the atomic system
to relax toward equilibrium state followed by the density operator
$\hat{\rho}_{3}=\hat{\rho}_{th}\left(\omega_{2},\beta_{c}\right)$.
Eigenvalues of the working medium depend only on the amplitude of
the external field, which are kept fixed at $E_{\pm}\left(\omega_{2}\right)$
. According to Eqs. (1) and (2), the atom exchanges energy with the
cold bath in the form of heat transfer and no work is performed by
the magnetic field. The amount of the heat transfer between the atom
and the cold bath is represented by

\textcolor{black}{
\begin{align}
Q_{c} & =-\frac{\hbar}{2}\omega_{2}\tanh\left(\frac{1}{2}\beta_{c}\hbar\omega_{2}\right)\nonumber \\
 & -Tr\left(\hat{H}\left(\omega_{2},0,\tau_{1}\right)\hat{\rho}^{\ast}\left(\omega_{2},\omega_{1},\alpha,\beta_{h},\tau_{1}\right)\right).
\end{align}
}

At stage III (3-4), a process of thermodynamic adiabatic expansion
would be carried out by isolating the atom from the cold bath and
making the magnetic field whirl around in the duration between $t=0$
to $t=\tau_{2}$. The vector of the magnetic field transforms from
$\boldsymbol{B}\left(2,0,0\right)$ {[}red arrow in Fig. 1(c){]} to
$\boldsymbol{B}\left(1,\alpha,t\right)$ {[}green arrow in Fig. 1(c){]}.
The Larmor frequency returns back to $\omega_{1}$ again, allowing
the density operator to unitarily evolve to $\hat{\rho}_{4}=\hat{\rho}\left(\omega_{1},\omega_{2},\alpha,\beta_{c},\tau_{2}\right)$.
\textcolor{black}{In a similar way, we move the magnetic field in
z direction instantly at time $\tau_{2}$, followed by $\boldsymbol{B}\left(1,0,\tau_{2}\right)$.
}With Eq. (15), the work done by the atom follows as\textcolor{black}{
\begin{equation}
W_{2}=W\left(\omega_{1},\omega_{2},\alpha,\beta_{c},\tau_{2}\right).
\end{equation}
}

At stage IV (4-1), since the Hamiltonian and the eigenenergies $E_{\pm}\left(\omega_{1}\right)$
are independent of time, a quantum isochoric evolution takes place
without any work perform. The atom develops into the original canonical
state $\hat{\rho}_{1}$ via thermalization with the hot bath at temperature
$T_{h}$, which makes the heat engine operate automatically in a cyclic
manner. At the end of this stage, the heat absorbed by the atom is
written as\textcolor{black}{
\begin{align}
Q_{h} & =-\frac{\hbar}{2}\omega_{1}\tanh\left(\frac{1}{2}\beta_{h}\hbar\omega_{1}\right)\nonumber \\
 & -Tr\left(\hat{H}\left(\omega_{1},0,\tau_{2}\right)\hat{\rho}^{\ast}\left(\omega_{1},\omega_{2},\alpha,\beta_{c},\tau_{2}\right)\right).
\end{align}
}

After completing one cycle, the total energy contained within the
atom always returns to its initial value. The net work of the cycle
turns out to be\textcolor{black}{
\begin{equation}
W=W_{1}+W_{2}=W_{L}+W_{S}.
\end{equation}
The second equality means that the work has been partitioned into
two distinct parts according to the separation in Eq. (15), where
$W_{L(S)}=W_{L(S)}\left(\omega_{2},\omega_{1},\alpha,\beta_{h},\tau_{1}\right)+W_{L(S)}\left(\omega_{1},\omega_{2},\alpha,\beta_{c},\tau_{2}\right).$}
For purposes of extracting work from the quantum heat engine, it is
necessary that $W<0$. Using Eqs. (17)-(21), we obtain the expression
of the efficiency as
\begin{equation}
\eta=\frac{-W}{Q_{h}}=1+\frac{Q_{c}}{Q_{h}}.
\end{equation}

\section{RESULTS AND DISCUSSION}

To build the complete descriptions of a quantum system in thermodynamic
processes, it is important to explain how to differentiate between
the quantum and thermodynamic adiabatic processes. For a quantum adiabatic
process, if a system starts from an eigenstate of the initial Hamiltonian
and the gaps among the eigenvalues exist, it will remain in the corresponding
instantaneous eigenstate of the final Hamiltonian when the perturbation
acting on it remains sufficiently slow \citep{key-41,key-42}, i.e.,
$\tau=\left|\hbar\left\langle m\right|\frac{\partial\hat{H}}{\partial t}\left|n\right\rangle /\left(E_{n}-E_{m}\right)^{2}\right|\ll1\left(n\neq m\right)$.
However, the first law of thermodynamics requires that a thermodynamic
adiabatic process occurs with the rate of heat transfer $\dot{Q}=0$,
relating the changes in internal energy only to the work done. The
thermodynamic adiabatic process does not require the quantum adiabatic
approximation to be satisfied \citep{key-11,key-43}. 

In the case of a zero rotation angle $\alpha=0$, the atom-field interaction
Hamiltonian becomes independent of time followed by $\frac{\partial\hat{H}\left(\omega_{j},\alpha,t\right)}{\partial t}=0$
{[}Eq. (5){]}. The density operators of the four terminal states of
the cycle {[}Fig. 1(a){]} fulfill the relations $\hat{\rho}_{2}=\hat{\rho}_{1}=\hat{\rho}_{th}\left(\omega_{1},\beta_{h}\right)$
and $\hat{\rho}_{4}=\hat{\rho}_{3}=\hat{\rho}_{th}\left(\omega_{2},\beta_{c}\right)$
regardless of the time scales. The occupation probability of each
instantaneous state remains unchanged during the transition from state
1 (3) to 2 (4), quantifying the applicability of quantum adiabatic
approximations. The heat absorbed from the hot bath and the net work
of the cycle could be, respectively, simplified as $Q_{h}=\frac{\hbar}{2}\omega_{1}\left[\tanh\left(\frac{1}{2}\beta_{c}\hbar\omega_{2}\right)-\tanh\left(\frac{1}{2}\beta_{h}\hbar\omega_{1}\right)\right]$
and $W=\frac{\hbar}{2}\left(\omega_{1}-\omega_{2}\right)\left[\tanh\left(\frac{1}{2}\beta_{h}\hbar\omega_{1}\right)-\tanh\left(\frac{1}{2}\beta_{c}\hbar\omega_{2}\right)\right]$.
For the heat engine operating at the quantum adiabatic limit, $\omega_{2}/\omega_{1}>\beta_{h}/\beta_{c}$
is certainly a necessary condition to create useful work from the
thermal energy. Meanwhile, the efficiency $\eta_{O}=1-\omega_{2}/\omega_{1}<1-\beta_{h}/\beta_{c}$,
which is automatically less than the Carnot efficiency. These results
appear consistent with prior researches based on two-level systems
\citep{key-6,key-44}. 

\begin{figure}
\includegraphics[scale=0.2]{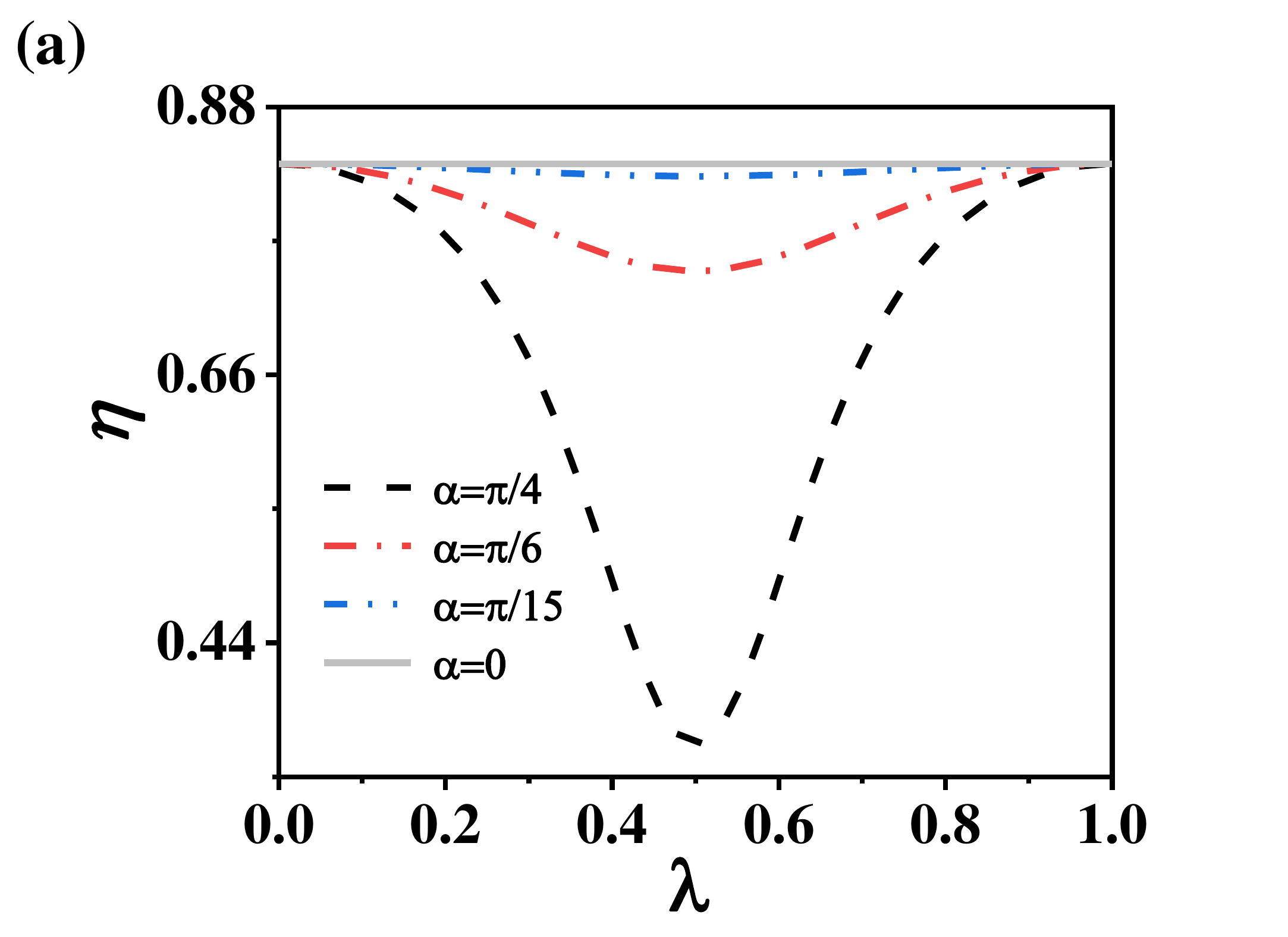}\includegraphics[scale=0.2]{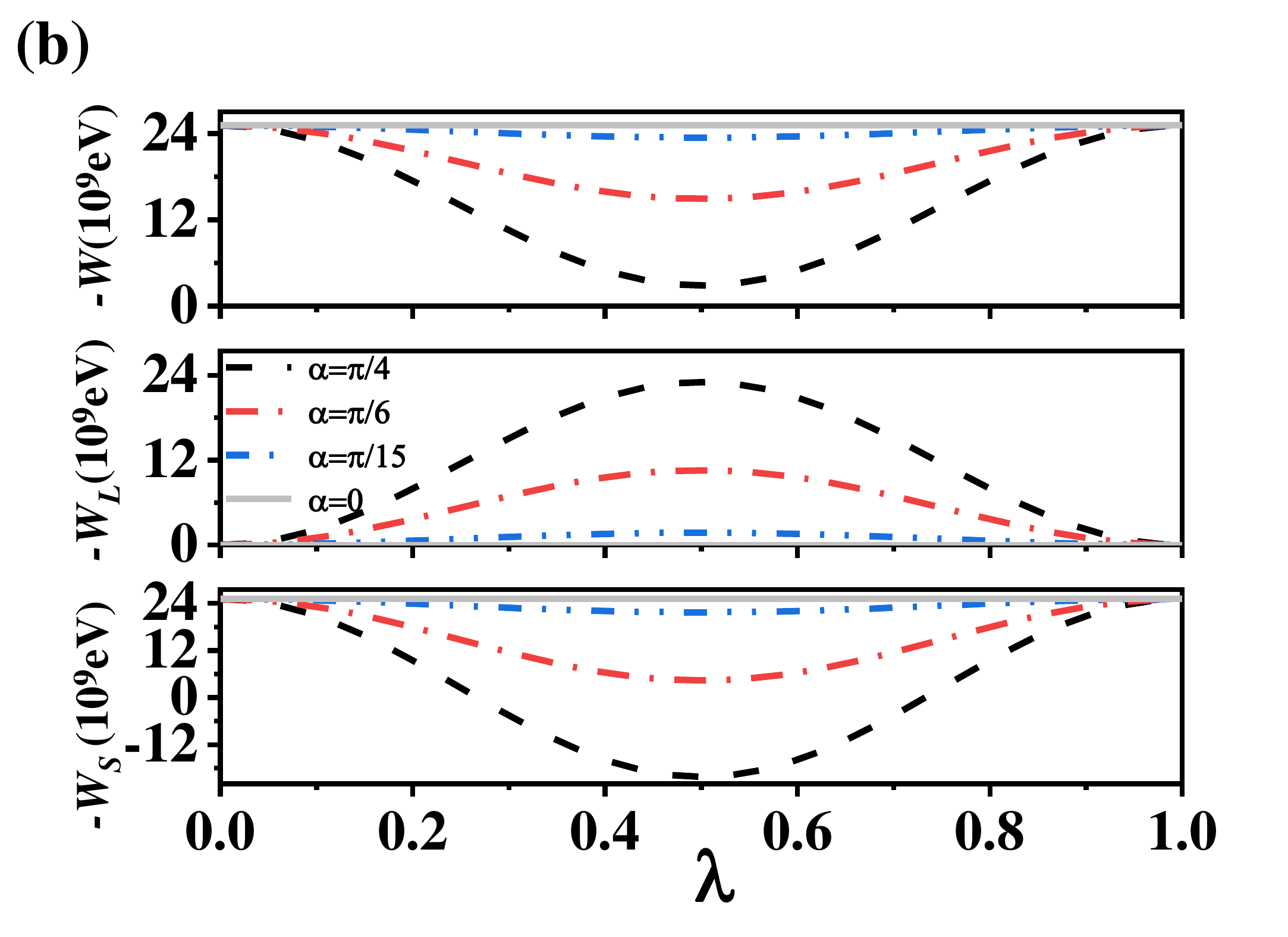}

\includegraphics[scale=0.2]{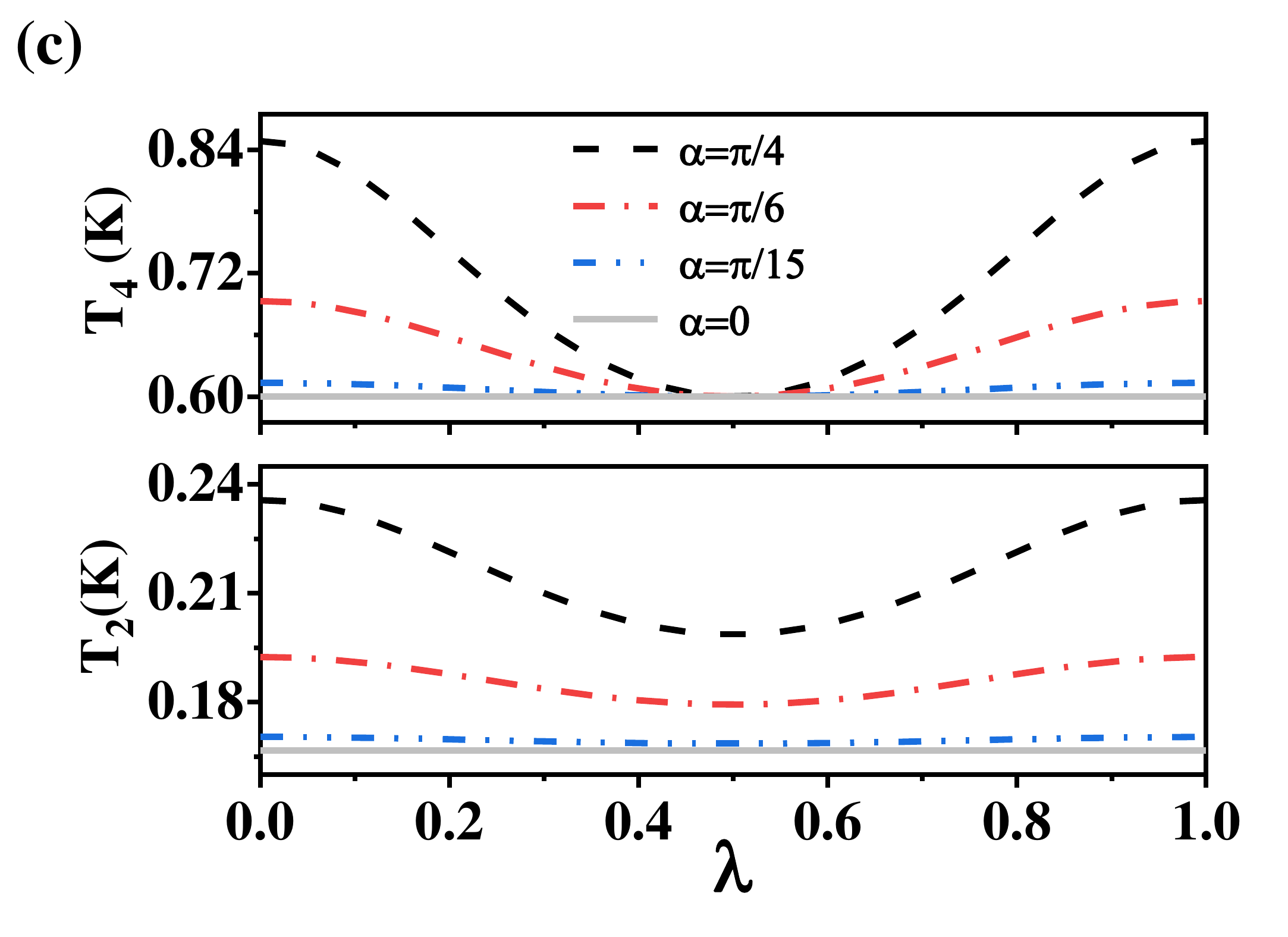}\includegraphics[scale=0.2]{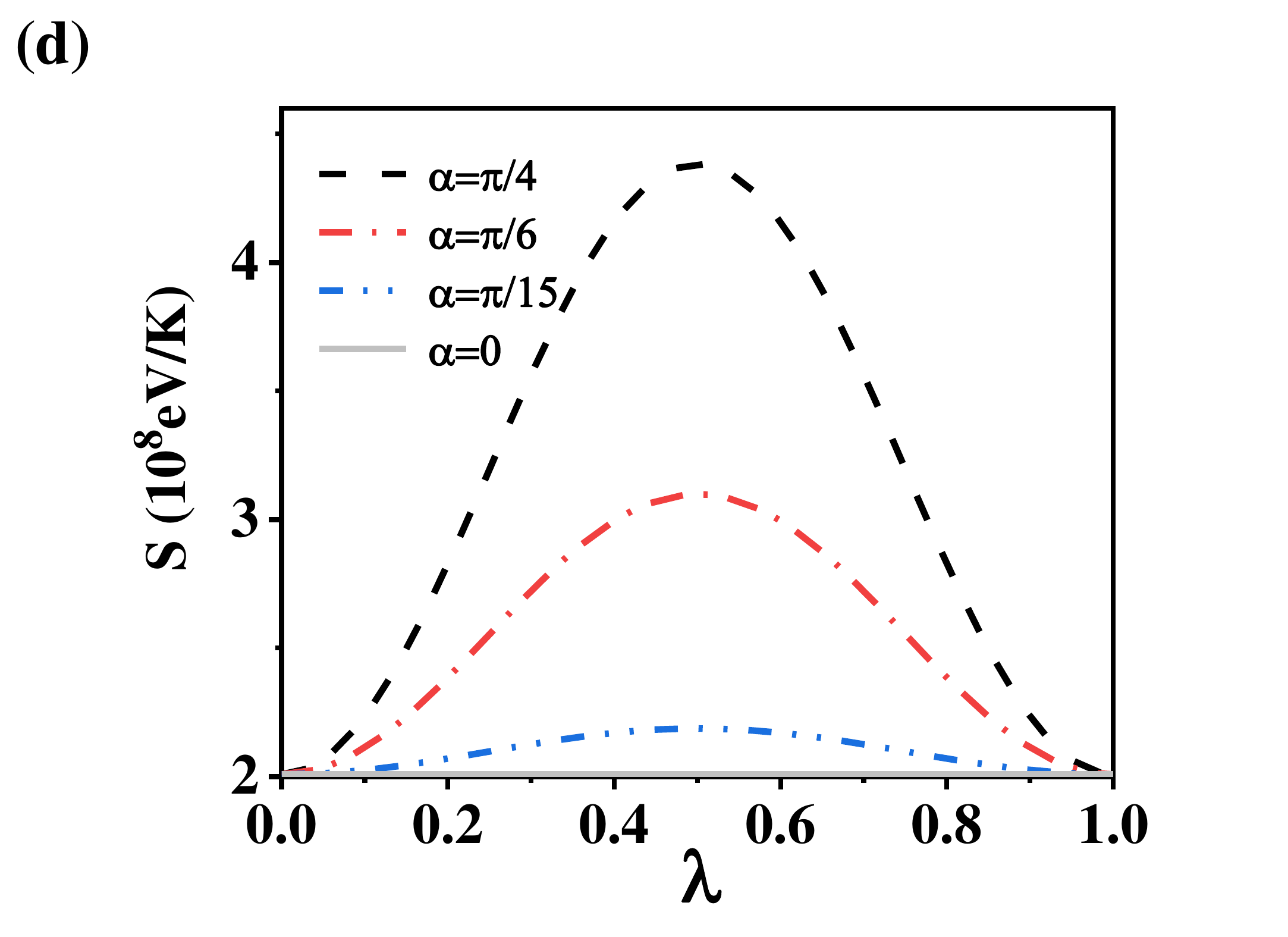}

\includegraphics[scale=0.2]{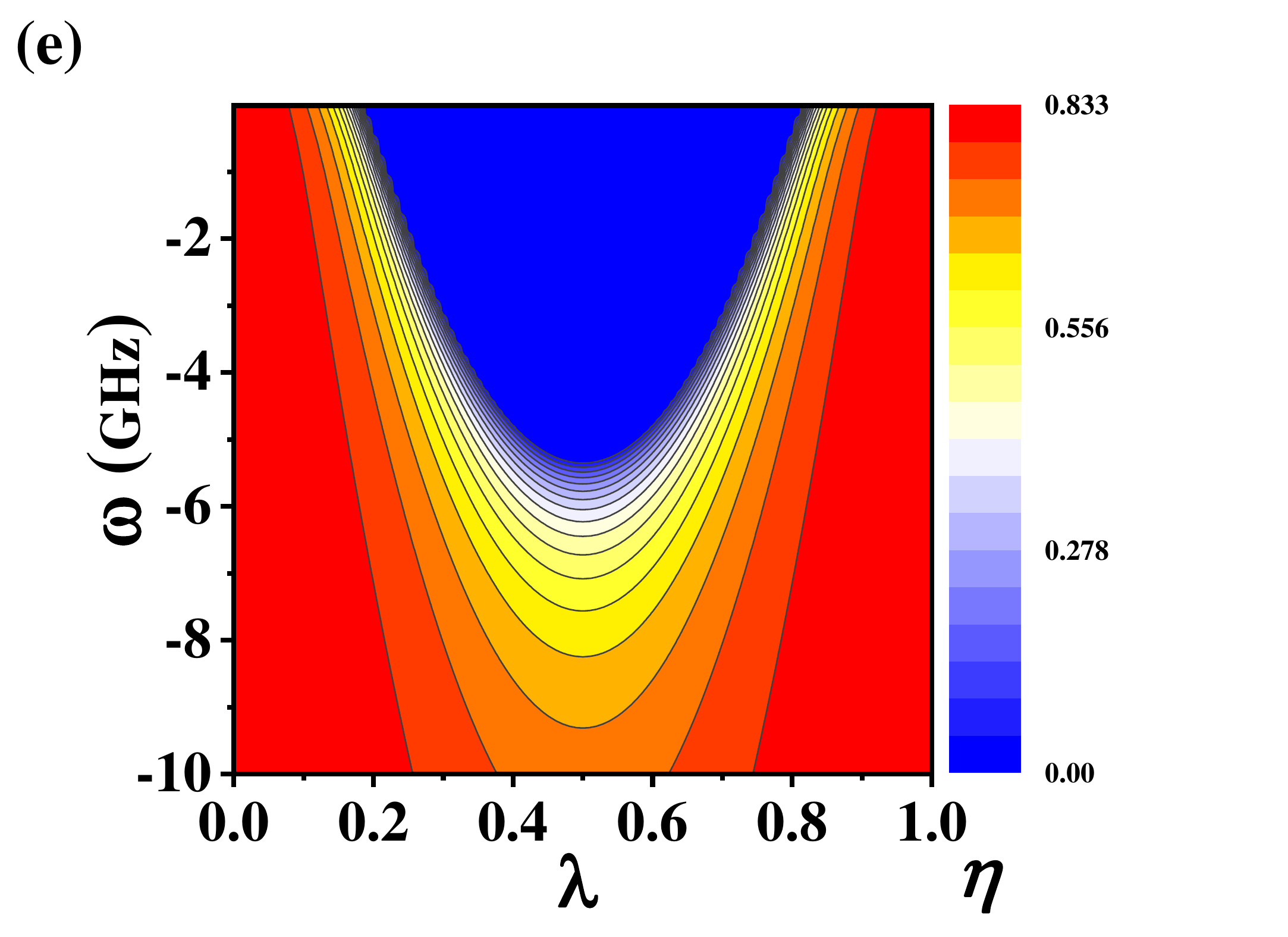}\includegraphics[scale=0.2]{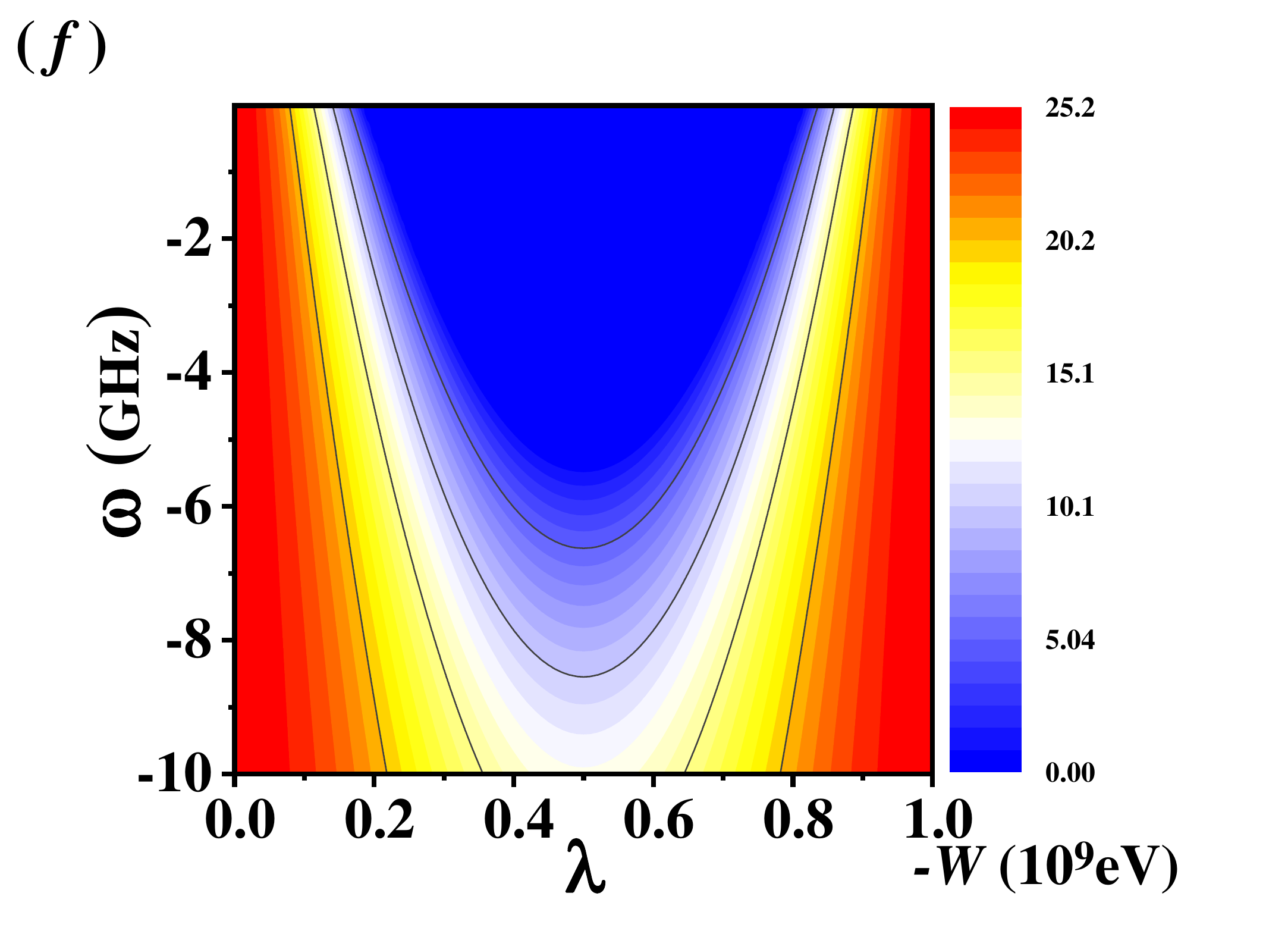}\caption{(a) The efficiency $\eta$, (b) work output $-W$, $-W_{L}$, and
$-W_{S}$, (c) effective temperatures $T_{2}$ and $T_{4}$, and (d)
entropy generation of the heat engine varying with the dimensionless
time parameter $\lambda$ at $\omega=-6GHz$ and the inclination angles
$\alpha=0$ (solid line, gray), $\pi/15$ (dash-double-dotted line,
blue), $\pi/6$ (dash-dotted line, red), and $\pi/4$ (dash line,
black). The contour plots of the (e) efficiency and (f) work output
as functions of $\lambda$ and the angular velocity $\omega$ at $\alpha=\pi/4$.
The parameters $T_{h}=1K$, $T_{c}=0.1K$, $\omega_{1}=6GHz$, and
$\omega_{2}=1GHz$.}
\end{figure}

\textcolor{black}{An investigation into the role of time-dependent
protocol will focus on the numerical simulation. The efficiency $\eta$
and work output $-W$ exhibit periodicity with respect to the time
spans of the thermodynamic adiabatic processes $\tau_{1}$ and $\tau_{2}$,
which repeat over intervals of $2\pi/\varOmega_{1}$ and $2\pi/\varOmega_{2}$.
By defining a dimensionless time parameter $\lambda=\frac{\tau_{1}\varOmega_{1}}{2\pi}=\frac{\tau_{2}\varOmega_{2}}{2\pi}$,
Figs. 2 (a) and (b) show the curves of the efficiency $\eta$ and
the work output $-W$ as functions of $\lambda$ at different values
of $\alpha$. When the angle between the z axis and the vector of
the field $\boldsymbol{B}\left(j,\alpha,t\right)$ goes to zero ,
i.e., $\alpha=0$, the efficiency reaches the Otto efficiency limit
$\eta_{O}$ under a transitionless driving {[}solid line in Fig. 2(a){]}.
The efficiency converges to the same result at $\lambda=0$ or 1.
This condition is equivalent to say that if the times for the adiabatic
driving $\tau_{1}$ and $\tau_{2}$ become finite and are set to be
integer multiples of $2\pi$, the heat engine will be operated in
the quantum adiabatic regime as well. The work output $-W$ compromises
the coherence work output $-W_{L}$ from the Larmor precession and
the work output $-W_{S}$ due to the sudden change of field. $-W_{L}$
increases as $\alpha$ increases from $0$ to $\pi/4$, but $-W_{S}$
may turn negative {[}dash line in Fig. 2(b){]}. Under these circumstances,
the coherence work facilitates the normal operation of the quantum
heat engine. From Figs. 2 (a) and (b), we conclude that the efficiency
and net work output in the quantum adiabatic regime impose a theoretical
limit of the heat engine with thermodynamic adiabatic processes. The
changes of the probabilities of different eigenstates due to the non-ideal
adiabatic processes ($\hat{\rho}_{2}\neq\hat{\rho}_{1}$ and $\hat{\rho}_{4}\neq\hat{\rho}_{3}$)
occur in association with a supernumerary entropy production, increasing
the irreversibility of the complete cycle. }

By assuming that the ratio of the two occupation probabilities satisfies
the Boltzmann distribution, the effective temperatures of the atom
$T_{2}$ and $T_{4}$ at state 2 and 4 are, respectively, depending
on the relations $\frac{\left\langle \chi_{+}\left(\alpha,\tau_{1}\right)\right|\hat{\rho}_{2}\left|\chi_{+}\left(\alpha,\tau_{1}\right)\right\rangle }{\left\langle \chi_{-}\left(\alpha,\tau_{1}\right)\right|\hat{\rho}_{2}\left|\chi_{-}\left(\alpha,\tau_{1}\right)\right\rangle }=e^{-\frac{E_{+}\left(\omega_{2}\right)-E_{-}\left(\omega_{2}\right)}{kT_{2}}}$
and $\frac{\left\langle \chi_{+}\left(\alpha,\tau_{2}\right)\right|\hat{\rho}_{4}\left|\chi_{+}\left(\alpha,\tau_{2}\right)\right\rangle }{\left\langle \chi_{-}\left(\alpha,\tau_{2}\right)\right|\hat{\rho}_{4}\left|\chi_{-}\left(\alpha,\tau_{2}\right)\right\rangle }=e^{-\frac{E_{+}\left(\omega_{1}\right)-E_{-}\left(\omega_{1}\right)}{kT_{4}}}$\citep{key-17}.
Figure 2 (c) demonstrates that the atom at state 2 is warmer than
the cold bath ($T_{2}>T_{c}$) and give up its energy to the cold
bath. Stage IV starts at a temperature smaller than that of the hot
bath ($T_{4}<T_{h}$) . As a result, the heat engine is taking a quantity
of heat energy from the hot bath until it reaches the equilibrium
state. During a closed cycle, the atom returns to its original thermal
state. The entropy generation per cycle $S=-\frac{Q_{h}}{T_{h}}-\frac{Q_{c}}{T_{c}}$.\textcolor{red}{{}
}\textcolor{black}{The entropy generation always remains positive
without violating the second law of thermodynamics. Particularly,
the quantum adiabatic regimes at $\alpha=0$ and $\lambda=0\left(1\right)$
allow a lower limit of $S$ to be obtained {[}Fig. 2(d){]}. }

\textcolor{black}{Under a time-dependent adiabatic evolution, the
efficiency $\eta$ and work output $-W$ can be enhanced by modulating
the dimensionless time parameter $\lambda$ and the angular velocity
$\omega$ {[}Figs. 2 (e) and (f){]}. For purposes of generating positive
work to the environment regardless of the time spent in the adiabatic
processes, the magnetic field should whirl around quickly. In the
limit of $\omega\rightarrow\infty$ , $\underset{\omega\rightarrow\infty}{\lim}\hat{U}\left(\omega_{j},\alpha,t\right)=1$.
As a result, the rapidly changing conditions prevent the system from
adapting its configuration during the process, and the probabilities
of the state remain unchanged. The efficiency is found close to the
quantum adiabatic limit $\eta_{O}$ again.}\textcolor{red}{{} }Based
on the division of heat and work in thermodynamic processes with quantum
coherence, one can conveniently design an efficient quantum Otto heat
engine concerning the time-dependent control.

\section{CONCLUSIONS}

A simple model of the QOHE with a time-dependent adiabatic process
is constructed in the frame of a spin driven by the rotating magnetic
field. On the basis of the first law of thermodynamics premeditating
the elements of quantum coherence, the work function relating to adiabatic
evolution from an arbitrary equilibrium state is obtained. When the
quantum cycle undergoes irreversible adiabatic processes, coherence
induced population transition ensures that the heat engine can work
properly. The efficiency and net work output at the quantum adiabatic
speed limit set a upper bound for a QOHE under time-dependent control.
The proposed model offers possible schemes to implement quantum cycles
by manipulating a single nuclear spin via a sequence of suitable pulses
and reconstructing the quantum state through the quantum state tomography.
\begin{acknowledgments}
This work has been supported by the National Natural Science Foundation
(Grant No. 11805159), the Fundamental Research Fund for the Central
Universities (No. 20720180011), and the Natural Science Foundation
of Fujian Province of China (No. 2019J05003). 
\end{acknowledgments}


\begin{thebibliography}{99}
\bibitem{key-1}\textcolor{black}{R. Dann, A. Tobalina, and R. Kosloff,
Phys. Rev. Lett. 122, 250402 (2019).}

\textcolor{black}{\bibitem{key-2}W. Niedenzu, V. Mukherjee, A. Ghosh,
A. G. Kofman, and G. Kurizki, Nat. Commun. 9, 165 (2018). }

\textcolor{black}{\bibitem[3]{key-3}R. Alicki, J. Phys. A: Math.
Gen. }\textbf{\textcolor{black}{12}}\textcolor{black}{, L103-L107
(1979).  }

\textcolor{black}{\bibitem[4]{key-4}F. Binder, L. A. Correa, C. Gogolin,
J. Anders, G. Adesso, Thermodynamics in the Quantum Regime-Fundamental
Aspects and New Directions, Switzerland, Springer, (2018).}

\textcolor{black}{\bibitem[5]{key-5}R. Kosloff, J. Chem. Phys. 80,
1625-1631 (1984). }

\textcolor{black}{\bibitem[6]{key-6}E. Geva and R. Kosloff, J. Chem.
Phys. 96, 3054-3067 (1992). }

\textcolor{black}{\bibitem[7]{key-7}Y. Rezek and R. Kosloff, New.
J. Phys. 8, 83 (2006). }

\textcolor{black}{\bibitem[8]{key-8}R. Kosloff, J. Chem. Phys. 150,
204105 (2019). }

\textcolor{black}{\bibitem[9]{key-9}E. Boukobza and D. J. Tannor,
Phys. Rev. Lett. 98, 240601 (2007). }

\textcolor{black}{\bibitem[10]{key-10}E. Boukobza and D. J. Tannor,
Phys. Rev. A 74, 063823 (2006). }

\textcolor{black}{\bibitem[11]{key-11}S. Su, J. Chen, Y. Ma, J. Chen,
and C. Sun, Chin. Phys. B 27, 060502 (2018). }

\textcolor{black}{\bibitem[12]{key-12}K. Brandner, M. Bauer, and
U. Seifert, Phys. Rev. Lett. 119, 170602 (2017). }

\textcolor{black}{\bibitem{key-13}J. Klatzow, J. N. Becker, P. M.
Ledingham, C. Weinzetl, K. T. Kaczmarek, D. J. Saunders, J. Nunn,
I. A. Walmsley, R. Uzdin, and E. Poem, Phys. Rev. Lett. 122, 110601
(2019). }

\textcolor{black}{\bibitem[14]{key-14}A. Ronzani, B. Karimi, J. Senior,
Y. Chang, J. T. Peltonen, C. Chen, and J. P. Pekola, Nat. Phys. 14,
991\textendash 995 (2018). }

\textcolor{black}{\bibitem[15]{key-15}B. Karimi and J. P. Pekola,
Phys. Rev. B 94, 184503 (2016). }

\textcolor{black}{\bibitem[16]{key-16}J. P. Pekola, Nat. Phys. 11,
118\textendash 123 (2015). }

\textcolor{black}{\bibitem[17]{key-17}R. J. de Assis, T. M. de Mendonça,
C. J. Villas-Boas, A. M. de Souza, R. S. Sarthour, I. S. Oliveira,
and N. G. de Almeida, Phys. Rev. Lett. 122, 240602 (2019). }

\textcolor{black}{\bibitem[18]{key-18}T.\LyXThinSpace B. Batalhão,
A.\LyXThinSpace M. Souza, L. Mazzola, R. Auccaise, R.\LyXThinSpace S.
Sarthour, I.\LyXThinSpace S. Oliveira, J. Goold, G. De Chiara, M.
Paternostro, and R.\LyXThinSpace M. Serra, Phys. Rev. Lett. 113, 140601
(2014).}

\textcolor{black}{\bibitem[19]{key-19}M. O. Scully, M. S. Zubairy,
G. S. Agarwal, H. Walther, Science, 299, 862-864 (2003).  }

\textcolor{black}{\bibitem[20]{key-20}H. T. Quan, P. Zhang, and C.
P. Sun, Phys. Rev. E 73, 036122 (2006).  }

\textcolor{black}{\bibitem[21]{key-21}M. T. Mitchison, M. P. Woods,
J. Prior, and M. Huber, New J. Phys. 17, 115013 (2015).}

\textcolor{black}{\bibitem[22]{key-22}J. Roßnagel, O. Abah, F. Schmidt-Kaler,
K. Singer, and E. Lutz, Phys. Rev. Lett. 112, 030602 (2014).  }

\textcolor{black}{\bibitem[23]{key-23}J. Klaers, S. Faelt, A. Imamoglu,
and E. Togan, Phys. Rev. X 7, 031044 (2017).  }

\textcolor{black}{\bibitem[24]{key-24}X. L. Huang, T. Wang, and X.
X. Yi, Phys. Rev. E 86, 051105 (2012). }

\textcolor{black}{\bibitem[25]{key-25}K. Zhang, F. Bariani, and P.
Meystre, Phys. Rev. Lett. 112, 150602 (2014). }

\textcolor{black}{\bibitem[26]{key-26}K. Zhang, F. Bariani, and P.
Meystre, Phys. Rev. A 90, 023819 (2014). }

\textcolor{black}{\bibitem[27]{key-27}A. Mari, A. Farace, and V.
Giovannetti, J. Phys. B: At. Mol. Opt. Phys. 48, 175501 (2015). }

\textcolor{black}{\bibitem[28]{key-28}G. Verley, M. Esposito, T.
Willaert, and C. V. den Broeck, Nat. Commun. 5, 4721 (2014). }

\textcolor{black}{\bibitem[29]{key-29}M. Campisi, J. Phys. A: Math.
Theor. 47, 245001 (2014). }

\textcolor{black}{\bibitem[30]{key-30}P. Pietzonka and U. Seifert,
Phys. Rev. Lett. 120, 190602 (2018). }

\textcolor{black}{\bibitem[31]{key-31}M. Bauer, K. Brandner, and
U. Seifert, Phys. Rev. E 93, 042112 (2016). }

\textcolor{black}{\bibitem[32]{key-32}H. Wang, J. He, and J. Wang,
Phys. Rev. E 96, 012152 (2017). }

\textcolor{black}{\bibitem[33]{key-33}Q. Liu, J. He, Y. Ma, and J.
Wang, Phys. Rev. E 100, 012105 (2019). }

\textcolor{black}{\bibitem[34]{key-34}C. Ou and S. Abe, Europhys.
Lett. 113, 40009 (2016). }

\textcolor{black}{\bibitem[35]{key-35}J. von Neumann, Mathematical
foundations of quantum mechanics, Princeton, Princeton University
Press (1955).}

\textcolor{black}{\bibitem[36]{key-36}T. D. Kieu, Phys. Rev. Lett.
93, 140403 (2004). }

\bibitem[37]{key-37}D. J. Griffiths, Introduction to quantum mechanics,
2nd Ed., New Jersey, Prentice Hall (2005).

\bibitem[38]{key-38}\textcolor{black}{Y. Ma, S. Su, and C. Sun, Phys.
Rev. E 96, 022143 (2017). }

\bibitem[39]{key-39}H. T. Quan, Phys. Rev. E 79, 041129 (2009). 

\bibitem[40]{key-40}G. A. Barrios, F. Albarrán-Arriagada, F. A. Cárdenas-López,
G. Romero, and J. C. Retamal, Phys. Rev. A 96, 052119 (2017). 

\bibitem[41]{key-41}M. V. Berry\textcolor{black}{, J. Phys. A: Math.
Theor. 42, 365303 (2009). }

\textcolor{black}{\bibitem[42]{key-42}C. P. Sun, J. Phys. A: Math.
Gen. 21, 1595-1599 (1988). }

\textcolor{black}{\bibitem[43]{key-43}H. T. Quan, Y. Liu, C. P. Sun,
and F. Nori, Phys. Rev. E 76, 031105 (2007). }

\textcolor{black}{\bibitem[44]{key-44}J. He, J. Chen, and B. Hua,
Phys. Rev. E 65, 036145 (2002).}
\end{thebibliography}
\end{document}